\documentclass[aps,twocolumn,prb,superscriptaddress]{revtex4}

\usepackage{graphicx}
\begin{document}
\title{Possible singlet to triplet pairing transition in Na$_x$CoO$_2\, \cdot\, y$H$_2$O }
\author{M. M. Ma{\'s}ka}
\author{M. Mierzejewski}
\affiliation{Department of Theoretical Physics, Institute of Physics, 40--007 Katowice, Poland}
\author{B. Andrzejewski}
\affiliation{Institute of Molecular Physics, PAS, Smoluchowskiego 17, 60--179 Pozna{\'n}, Poland}
\author{M. L. Foo}
\affiliation{Department of Chemistry,  Princeton University, Princeton,
New Jersey 08545}
\author{R. J. Cava}
\affiliation{Department of Chemistry, Princeton University, Princeton,
New Jersey 08545}
\author{T. Klimczuk}
\affiliation{Department of Chemistry,  Princeton University, Princeton,
New Jersey 08545}
\affiliation{Faculty of Applied Physics and Mathematics, Gdansk University of Technology, Narutowicza 11/12, 
80-952 Gdansk, Poland}
\begin{abstract}
We present precise measurements of the upper critical field ($H_{c2}$) in the recently
discovered cobalt oxide superconductor. We have found that the critical field
has an unusual temperature dependence; namely, there is an abrupt change of
the slope of $H_{c2}(T)$ in a weak field regime. 
In order to explain this result  we have derived 
and solved Gor'kov equations on a triangular lattice. Our experimental
results may be interpreted in terms of the field--induced transition
from singlet to triplet superconductivity. 
\end{abstract}
\pacs x 74.25.Op, 74.20.Rp
\maketitle
\section{Introduction}
The recent discovery of superconductivity in 
${\rm Na}_x {\rm Co O_2} \cdot y {\rm H_20}$,\cite{takada}
may provide a unique insight into the mechanisms, which
determine superconducting properties of transition metal oxides.
Although the superconducting transition temperature, $T_c$, is much lower
than $T_c$'s in cuprate superconductors, both system share many common features.
${\rm Co}$ oxide becomes superconducting after hydration that
significantly enhances the distance between ${\rm Co O_2}$ layers.
This suggests crucial importance of the dimensionality.
In particular, a quasi two--dimensional character of ${\rm Co}$ oxide  shows up in
the resistivity measurements. Above the transition temperature,
the in--plane resistivity is three orders of magnitude less than out--of--plane one.\cite{jin} 
Similarly to cuprates the ${\rm Co}$--based superconductor represents a strongly
correlated system. The strong correlations may be responsible for a nonmonotonic doping 
dependence of $T_c$. Namely, the critical temperature is maximal for 
a particular carrier concentration and decreases both 
for overdoped and underdoped materials.\cite{tomek}    

However, in contradistinction to cuprates,  ${\rm Co O_2}$ layers have a form of a triangular
lattice. This feature may be responsible for magnetic frustration and 
unconventional symmetry of the superconducting order parameter.
Investigations of the pairing symmetry with the help
of nuclear magnetic resonance (NMR)\cite{nmr} and nuclear quadrupole resonance (NQR)\cite{nqr}
lead to contradictory results. In particular, the presence of nodes in 
the superconducting gap remains an open problem. It is also unclear whether
superconductivity originates from singlet or triplet pairing. Theoretical
investigations do not lead to firm conclusions. It has been shown that the
resonating valence bond state (RVB) may be realized in the $t$--$J$ model
on a triangular lattice,\cite{koretsune} provided $t>0$. 
This may suggest RVB as a straightforward explanation 
of superconductivity in the ${\rm Co}$ oxides.\cite{baskaran}
However, it is interesting that in addition to singlet superconductivity, 
there is a region of triplet pairing in the phase diagram proposed in Ref. \cite{baskaran}. 
 Moreover, LDA 
calculations suggest that the ground state of the parent system
${\rm NaCo_2O_4 }$ may be ferromagnetic.\cite{singh} 
Recent density functional calculations carried out for
${\rm Na}_x{\rm Co_2O_4 }$ predict an itinerant ferromagnetic
state that, however, competes with a weaker antiferromagnetic
instability.\cite{singh1} Triplet superconductivity
has also been postulated on the basis of symmetry considerations
combined with analysis of experimental results.\cite{tanaka}

Therefore, the symmetry of the superconducting order
parameter  remains an open problem and both singlet and triplet pairings
should seriously be taken into account. In particular, it is possible
that singlet and triplet types of superconductivity compete with
each other. In such a case an external magnetic field may favor 
triplet pairing, due to the absence of the Zeeman pair breaking mechanism in this state.
This should be visible in the temperature dependence of the upper critical field, $H_{c2}$.
In order to verify this possibility we carry out precise measurements of $H_{c2}$.
The obtained results clearly indicate unconventional temperature dependence of $H_{c2}$, that cannot be described
within the Werthamer--Helfand--Hohenberg (WHH) theory.\cite{whh}
The experimental data are compared with theoretical results obtained from the solution
of the Gor'kov equations on a triangular lattice. These results 
may be interpreted in terms of a field--induced transition from singlet to 
triplet superconductivity and suggest that phase sensitive measurements to distinguish
this from other possible interpretations would be of great interest.

\section{experimental results}

The measurements have been carried out on  Na$_{0.3}$CoO$_2\, \cdot\, 1.3$H$_2$O.
Na$_{0.7}$CoO$_2$ (0.5g) was stirred in 20 ml of a 40x Br$_2$ solution
in acetonitrile at room temperature for 4 days ('1x' indicates that the amount
of Br$_2$ used is exactly the amount that would theoretically be needed
to remove all the Na from Na$_{0.7}$CoO$_2$). The product was washed
copiously with acetonitrile, followed by water and air--dried. After
air--drying, the product was kept in a sealed container with 100\%
relative humidity for 2 days to obtain the hydrated superconductor.

\begin{figure}
\vspace*{3mm}
\includegraphics[width=8cm]{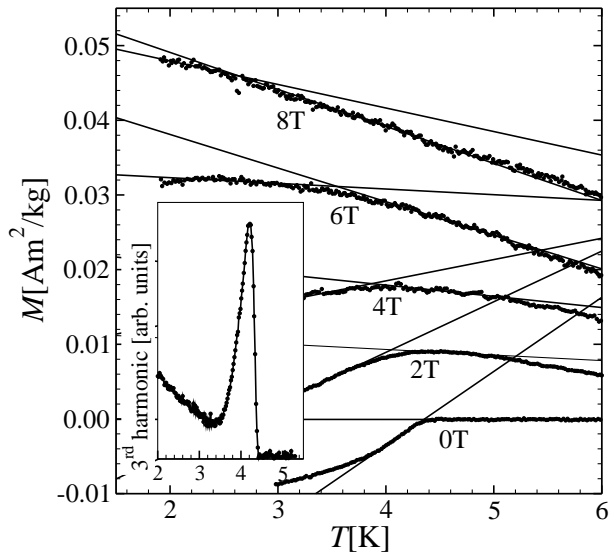}
\caption{Temperature dependence of mass magnetization $M(T)$ for various magnitudes 
of magnetic fields. For clarity of the figure the $M(T)$ curves are offset, 
except the one for 0T. The inset presents temperature dependence of the 3$^{\rm rd}$ 
harmonic susceptibility.
}
\label{chi}
\end{figure}

\begin{figure}
\vspace*{3mm}
\includegraphics[width=7.5cm]{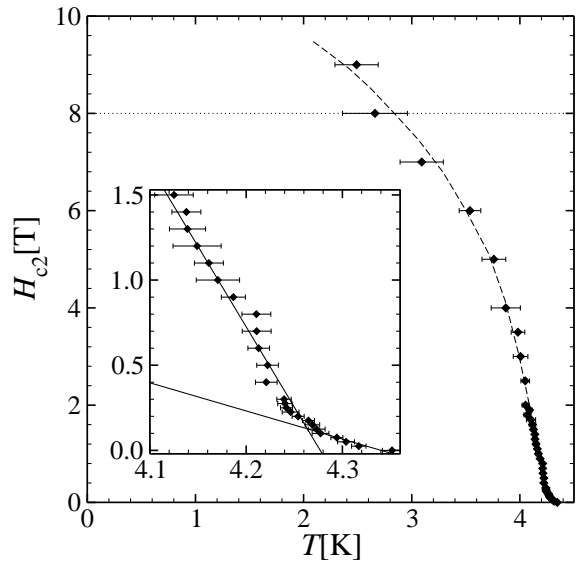}
\caption{
Experimental data for $H_{c2}(T)$. The horizontal line
shows the the Clogston--Chandrasekhar limit, whereas the line
connecting experimental points is only a guide for the eye. The insert shows
$H_{c2}(T)$ in the low field regime fitted by two lines.
}
\label{fit}
\end{figure}

All the magnetic measurements were performed using DC magnetometer/AC susceptometer 
MagLab 2000 System (Oxford Instruments Ltd.). 
There is only one superconducting phase transition in the sample and 
there is no coexistence of phases with different critical temperatures or 
critical fields. It has been well confirmed by a single peak 
in a temperature dependence of magnitude of the 3$^{\rm rd}$ harmonic AC susceptibility
(see the inset in Fig. \ref{chi}). 
For third harmonic measurements the 
AC magnetic field of frequency $f$=1kHz and amplitude $H_{\rm ac}=10^{-4}$T was 
applied. For the DC measurements we
%
%
have applied magnetic fields up to 9T. 
The temperature was stepped in the range about 3$\div$6K
in the case of low and moderate magnetic fields and 
in the range about 2$\div$5K for the highest fields. 
The size of the step was 20mK and the temperature was stabilized 
during each measurement with the accuracy 2mK. 
A set of typical $M(T$) curves recorded for applied fields of 
0, 2, 4, 6, and 8T is shown in Fig. \ref{chi}. 
Except the case of lowest magnetic fields the magnetization 
is positive in the whole temperature range. 
It is due to the domination of ferromagnetic and/or paramagnetic 
contributions in the total magnetization at higher fields. 
The superconducting transition manifests itself as a 
downturn in $M(T)$ at low magnetic fields, whereas at the higher fields 
only the change in the slope in $M(T)$ is observed. 
This enabled a simple determination of the critical temperature; namely,
$T_c$ was determined from the intersection of the two straight--lines that fit
relevant linear regimes (see Fig. \ref{chi}). The zero--field critical temperature determined in this way
is $T_c(0)= 4.345 \pm 0.015$~K.
The results of the measurements are presented in Fig. \ref{fit} in equivalent form as $H_{c2}(T)$. 

Close to $T_c(H=0)$  one can expect
that the Ginzburg--Landau theory gives accurate results and, 
therefore, temperature dependence of $H_{c2}$ should be linear.
In cuprate superconductors $H_{c2}$ shows unconventional 
temperature dependence.\cite{osofsky,mackenzie} 
Close to $T_c(H=0)$ $H_{c2}(T)$ is almost linear and, then,
the curvature smoothly increases with the decreasing temperature.
However, as can be inferred from Fig. ~\ref{fit},
this is not the case for ${\rm Na}_x {\rm Co O_2} \cdot y {\rm H_20}$.
For $1{\rm T} \alt H \alt 3$T 
the experimental data can be fitted very well by a linear function. 
However, such a fit deviates from experimental points for weaker magnetic field.
Similar temperature dependence of $H_{c2}$ has been obtained, e.g., from 
the specific heat measurements.\cite{yang} 
For $H\alt 3T$ the experimental data presented in Fig. ~\ref{fit},
as well as those reported in Ref. \cite{yang}, can be fitted by two linear functions.  
In our case they are: $H_{c2}(T)= 7.4-1.7T$ in the weak field regime 
and $H_{c2}(T)=40-9.4T$ for stronger magnetic field.  
Using the WHH formula,\cite{whh} 
$H_{c2}(0) \simeq 0.7 \: T_c \left({\rm d} H_{c2} /{\rm d} T \right)|_{T_c} $, 
one estimates
corresponding values of $H_{c2}(0)$'s equal to 5.2T and 28T, respectively. 
Such a behavior may originate from competition between two superconducting order parameters
with close transition temperatures but different temperature dependences of $H_{c2}$.
Singlet and triplet order parameters are possible candidates due to the absence
of Zeeman pair breaking in the latter case. 
$H_{c2}(0)$ obtained from the WHH formula in the strong field regime is beyond the 
Clogston--Chandrasekhar (CC) limit. $H_{c2}(0)$'s reported in Ref. \cite{hc2hc1} and 
estimated from Ref. \cite{yang} are even higher.
Although, the extrapolated $H_{c2}(0)$ may be overestimated, our experimental data clearly show
that $H_{c2}$ exceeds the CC limit already for $T \simeq 0.6\:T_c$. The large slope
of $H_{c2}(T)$ suggests that even in the case of renormalization of the paramagnetic pair
breaking mechanism (e.g., similar
to that in the strong--coupling electron--phonon approach\cite{carbotte})
$H_{c2}(0)$ should be beyond the CC limit.
This speaks in favor of triplet superconductivity. 
On the other hand, $H_{c2}(0)$ estimated from the low field data does not exceed CC limit. 
Therefore, superconductivity in a weak magnetic field may originate from
the singlet pairing.    
In the following we show that this tempting interpretation of experimental
data remains in agreement with theoretical results obtained from the
numerical solution of the Gor'kov equations. 
Our fit neglects a positive curvature of $H_{c2}(T)$ that occurs for $H \alt 0.9$~T. 
At the end of this paper, we discuss possible origins of this feature.

\section{Theoretical approach to the upper critical field}

In order to calculate the upper critical field we consider a triangular
lattice immersed in a uniform perpendicular magnetic field:
\begin{eqnarray}
H&=&\sum_{\langle ij \rangle\sigma} t_{ij} e^{i\theta_{ij}}
c^\dagger_{i\sigma}
c_{j\sigma}
-\mu \sum_{i,\sigma} c^{\dagger}_{i\sigma}c_{i\sigma} \nonumber \\
&&-\:  g\mu_{B}H_{z}  \sum_{i}
\left(c^{\dagger}_{i\uparrow}c_{i\uparrow}
-c^{\dagger}_{i\downarrow}c_{i\downarrow} \right) 
\nonumber \\
&&+\:V^{\rm s}\:\:\sum_{\langle ij \rangle} \left(
\Delta_{ij} c^\dagger_{i\uparrow}c^\dagger_{j\downarrow}+ {\rm h.c.} \right) 
\nonumber \\
&&+\:V^{\rm t}\:\:\sum_{\langle ij \rangle}
\sum_{\sigma_1,\sigma_2=\uparrow\downarrow} \left(
\Delta_{ij}^{\sigma_1\sigma_2}c^\dagger_{i\sigma_1}c^\dagger_{j\sigma_2}+ {\rm h.c.} \right).
\label{hamil}
\end{eqnarray}
$t_{ij}$ is the hopping integral between the sites $i$ and $j$
in the absence of magnetic field and $\theta_{ij}$ is the Peierls phase
factor, responsible for the diamagnetic response of the system:
\begin{equation}
\theta_{ij}=\frac{2\pi}{\Phi_0}\int_i^j \vec{A}\cdot\vec{dl}, 
\label{Peierls}
\end{equation} 
where
${\Phi_0}=hc/e$ is the flux quantum. The chemical potential $\mu$ has been introduced
in order to control the carrier concentration.
In the Hamiltonian (\ref{hamil})
\begin{equation}
\Delta_{ij}=\langle c_{i\uparrow}c_{j\downarrow}
-c_{i\downarrow}c_{j\uparrow} \rangle
\end{equation}
and 
\begin{eqnarray}
\Delta^{\uparrow\downarrow}_{ij}&=&\langle c_{i\uparrow}c_{j\downarrow}
+c_{i\downarrow}c_{j\uparrow} \rangle, \\ 
\Delta^{\uparrow\uparrow}_{ij}&=&\langle c_{i\uparrow}c_{j\uparrow}
\rangle, \\
\Delta^{\downarrow\downarrow}_{ij}&=&\langle c_{i\downarrow}c_{j\downarrow}
\rangle  
\end{eqnarray}
 denote the pairing amplitudes in singlet and triplet channels, respectively.



In order to determine the upper critical field we follow the procedure 
introduced in Refs. \cite{mmmm2,mmmm1}. Namely, we diagonalize the kinetic 
part of the Hamiltonian [the first
term in Eq. (\ref{hamil})] by introducing a new set of fermionic operators. 
In the Landau gauge $\vec{A}=(-H_zy,0,0)$  this set of fermionic operators is 
determined by a one--dimensional eigenproblem known as the Harper equation. 
We restrict further considerations to the nearest--neighbor hopping, i.e.,
$t_{ij}=t$ for the neighboring sites $i,j$ and 0 otherwise.

\begin{figure}
\vspace*{3mm}
\includegraphics[width=5cm]{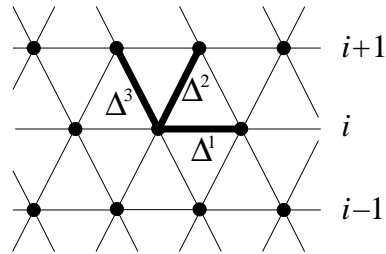}
\caption{Order parameters for superconductivity on a triangular lattice.
Such a set of order parameters is introduced in 
each pairing channel, i.e., for $\Delta,\ \Delta^{\uparrow\downarrow},\ \Delta^{\uparrow\uparrow},
\ \Delta^{\downarrow\downarrow}$.
}
\label{triangular}
\end{figure}

Using the solutions of the Harper equation we 
write down a self--consistent equation for the gap functions. 
Magnetic field breaks the translational symmetry and, therefore, the order parameters
are site dependent. However, in the chosen gauge they depend on the $y$ coordinate only.
As the superconductivity develops on the triangular lattice,
at each site we introduce three order parameters ($\Delta^1,\Delta^2,\Delta^3$)
 in each pairing channel,
i.e., for $\Delta,\ \Delta^{\uparrow\downarrow},\ \Delta^{\uparrow\uparrow},
\ \Delta^{\downarrow\downarrow}$ (see Fig. ~\ref{triangular}).
In order not to assume any particular pairing symmetry we consider 
these order parameters as independent quantities.
In the following we do not assume any particular orbital symmetry of the pair state.
However, independently of this symmetry all these order parameters vanish at $T_c$.
Therefore, the gap equation can be expressed with the help of three
vectors $\vec{\Delta}^{1,2,3}$, where $\vec{\Delta}^a=\left(
\Delta^a_1,\ \Delta^a_2,\ \ldots\right)$. The lower index enumerates 
rows of the lattice sites, whereas the upper one indicates the
direction of the bond, as depicted in Fig.~\ref{triangular}.
$H_{c2}$ is defined as a field, at which all components  of these vectors 
vanish. 
This can be determined from the linearized 
version of the gap equation that is of the following form:
\begin{equation}
\left(
\begin{array}{c}
\vec{\Delta}^1 \\
\vec{\Delta}^2 \\
\vec{\Delta}^3
\end{array}
\right)=\left(
\begin{array}{ccc}
& &  \\
& {\cal M} &  \\
& &  
\end{array}
\right)
\left(
\begin{array}{c}
\vec{\Delta}^1 \\
\vec{\Delta}^2 \\
\vec{\Delta}^3 
\end{array}
\right)
\label{matrix}
\end{equation}


For the sake of brevity we do not present an explicit form of ${\cal M}$.
This matrix can be expressed with the help of the Cooper pair susceptibility and eigenfunctions
obtained from the Harper equation.
\begin{figure}
\vspace*{3mm}
\includegraphics[width=8cm]{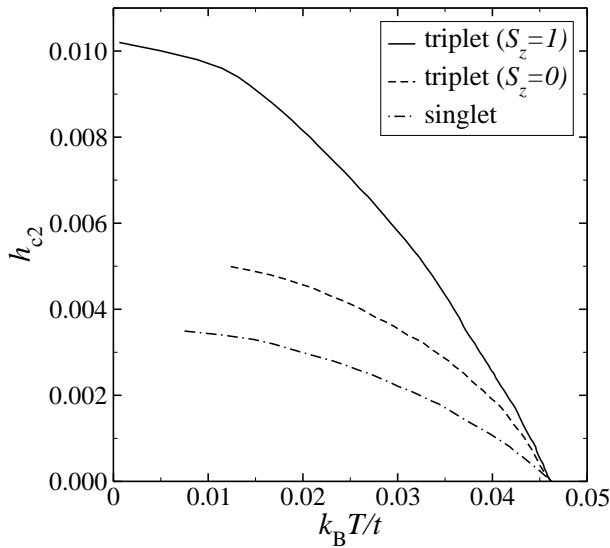}
\caption{Numerical results for $h_{c2}(T)$. Presented results have been
obtained for occupation number
$n=0.95$ and for $V^s=0.55t$, $V^t=0.75t$.}
\label{nasz1}
\end{figure}
\begin{figure}
\vspace*{3mm}
\includegraphics[width=8cm]{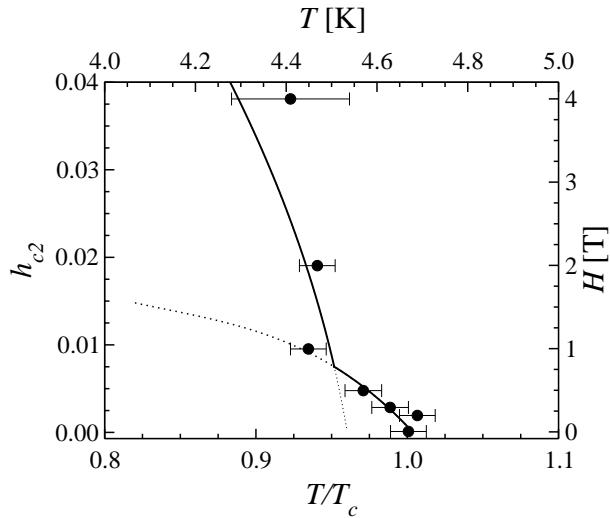}
\caption{
Numerical results for $h_{c2}(T)$. We have assumed the model parameters 
$n=0.67$ and $V^s=V^t=0.7t$. For these
model parameters $kT_c \simeq 0.2t$.
 The experimental points have been taken from Ref. \cite{yang}}
\label{nasz2}
\end{figure}

The temperature dependence of $H_{c2}$ has been obtained from Eq. (\ref{matrix}) for singlet
and triplet superconductivity. In the latter case we have investigated separately the paired states 
$\mid\downarrow\downarrow \rangle$,
$1/\sqrt{2}(\mid \uparrow\downarrow \rangle + \mid\downarrow \uparrow\rangle) $, and
$\mid \uparrow\uparrow \rangle$. We refer to these states by the corresponding
spin projection $S_z=-1,0,1$, respectively.
These states are affected by the
magnetic field in different ways due to the Zeeman coupling. In the case of equal--spin--pairing
this coupling is almost ineffective as it leads to a renormalization of the chemical potential only
$\mu \rightarrow \tilde{\mu}= \mu \pm  1/2\: g\mu_{B}H_{z} $.

Figures \ref{nasz1} and \ref{nasz2} show the numerical results
obtained for $150 \times 150$ cluster.
In particular, in Fig.  \ref{nasz2} we have chosen
the occupation number that is close to the experimentally determined
optimal doping.
We present a reduced magnetic field $h=2 \pi \Phi/\Phi_0$,
where $\Phi_0$ is the flux quantum and $\Phi$
is the magnetic flux through the lattice cell.
When the zero--field transition temperatures for 
singlet and triplet pairings are
of comparable magnitudes, the slope of $H_{c2}(T)$ is much larger for
triplet superconductivity. As a result, the triplet superconductivity
is characterized by much larger $H_{c2}(T=0)$.
It is interesting that
this feature remains valid also for triplet pairing with $S_z=0$
that is affected by the Zeeman pair breaking (see Fig. \ref{nasz1}).   
As expected, the highest value of the
upper critical field is obtained for the triplet equal--spin--pairing.
Comparing Figs. \ref{nasz1} and \ref{nasz2} 
one can see that the above statements on the slope of $H_{c2}(T)$
hold in a wide range
of the model parameters. However, due to limitation of the
cluster approach it is difficult to perform 
numerical calculations at very low temperatures.\cite{mmmm2}
From Figs. \ref{nasz2} and \ref{fit}  follows that 
experimental data are qualitatively reproduced when
$T_c(H=0)$ for triplet superconductivity is slightly less
than the transition temperature for the singlet one.
Then, sufficiently strong magnetic field leads to a transition
from singlet to triplet superconductivity that
shows up in a change of the slope of $H_{c2}(T)$.

The external magnetic field affects the relative 
phase of the order parameter in different directions
presented in Fig. \ref{triangular}. According to Eq. (\ref{Peierls}) this phase
can change from site to site and, therefore, it is 
impossible to determine globally the type of 
the symmetry of the energy gap. However, we have
found that for singlet pairing $T_c(H=0)$ is exactly
the same as transition temperature obtained for 
$d_1+id_2$ symmetry, according to the notation in Ref. \cite{baskaran}. On the other
hand, for the triplet pairing  $T_c(H=0)$ corresponds
to that for the $f$--wave pairing when $n \alt 1$ and
$p_x+ip_y$ symmetry for larger occupation number.
The exact position of the boundary between both
the triplet solutions depends on the pairing potential.

\section{Discussion and conclusions}

Our linear fit to the experimental data in the intermediate field
regime seems to be very accurate, strongly supporting
the triplet pairing.  Even stronger evidence 
comes from the presence of superconductivity
above the CC limit.
However, at weak field
there exists also other possibility:  $H_{c2}(T)$
for the field less then approx. 0.9T 
can be fitted by
a concave curve.  Similarity between 
${\rm Na}_x {\rm Co O_2} \cdot y {\rm H_20}$ 
and high--$T_c$ superconductors may suggest
a common mechanism that leads to the positive curvature
of $H_{c2}(T)$. This also speaks in favor of a singlet pairing in this regime.   
In such an approach the singlet--triplet transition takes place at slightly
higher field, $H \approx 0.9$ T.
From a theoretical point of view the upward curvature
of $H_{c2}(T)$ can occur for instance in:
extremely type II superconductors described by the
boson--fermion model,\cite{bosonfermion}
  the systems with a strong disorder sufficiently close
  to the metal--insulator transition,\cite{Kotliar-86}
  the disordered superconductors due to mesoscopic
  fluctuation,\cite{Spivak-95}
  Josephson tunneling between superconducting clusters,\cite{geshkenbein}
  in a mean--field--type theory of $H_{c2}$ with a strong
  spin--flip scattering,\cite{kresin}
  and due to a reduction of the diamagnetic pair--breaking
  in the stripe phase.\cite{last}
 Other theoretical approaches to this problem include, e.g.,
  the superconductivity with a mixed symmetry
  ($s+d$) order parameter\cite{Joynt-92} and
  Bose-Einstein condensation of charged bosons.\cite{alex}

$H_{c2}(T)$ obtained from the resistivity measurements\cite{sasaki} is lower
than presented here and, e.g., in Refs. \cite{hc2hc1,yang}. In particular, it is lower
than $H_{c2}(T)$ obtained from magnetization measurements on quasi--single
crystals.\cite{badica} This discrepancy remains unexplained.
One of possible explanations  is that it originates from the presence of lattice
defects, that for short coherence length superconductors form Josephson junctions.
These junctions affect the resistivity measurements much stronger than
the magnetization ones.


To conclude, we have measured the temperature 
dependence of the upper critical field in
\mbox{Na$_{0.3}$CoO$_2\, \cdot\, 1.3$H$_2$O} and we have found
an interesting feature;
namely, an abrupt change of slope of $H_{c2}(T)$ 
in a weak--field regime. This feature is in qualitative
agreement with results reported in Ref. \cite{yang}.
Moreover, such a bend in a weak field regime is visible also 
in other magnetization measurements,\cite{hc2hc1}
in specific heat measurements\cite{yang} 
as well as in resistivity measurements 
(see Fig. 4a in Ref. \cite{sasaki}.)
In order to explain the origin of such a behavior
we have solved Gor'kov equations on a triangular 
lattice for singlet and triplet types of pairing.
Our experimental results are consistent with a scenario
of competing singlet and triplet superconductivity.
Within such an approach magnetic field induces
a transition from singlet to triplet superconductivity
that shows up in a change of slope of $H_{c2}(T)$:
in a weak magnetic field the singlet pairing takes place and
sufficiently strong magnetic field drives the system into the triplet state. 
Recently, a field--induced transition between various types of 
singlet superconductivity has been proposed 
to take place in cuprates\cite{Krishana} (an occurrence of a minor 
$id_{xy}$ component of the order parameter).

\acknowledgments
B. A. acknowledges support from the Centre of Excellence for
Magnetic and Molecular Materials for Future Electronics within the EC Contract no. G5MA-CT-2002-04049;
M. Ma{\'s}ka acknowledges support from the Polish State Committee for Scientific Research, Grant No. 2 P03B 050 23.
T.K. acknowledges support from the Foundation for Polish Science. The work at Princeton is supported 
by the Department of Energy, grant DE--FG02--98--ER45706 and the NSF, grant DMR--0244254.

\end{document}